\documentclass[11pt]{article}

\usepackage{acl}
\usepackage{times}
\usepackage{latexsym}
\usepackage[T1]{fontenc}
\usepackage[utf8]{inputenc}
\usepackage{microtype}
\usepackage{inconsolata}
\usepackage{graphicx}
\usepackage{amsmath}
\usepackage{amssymb}
\usepackage{booktabs}
\usepackage{xcolor}
\usepackage[dvipsnames]{xcolor}
\usepackage[most]{tcolorbox}
\usepackage{colortbl}
\usepackage{array}
\usepackage{multirow}
\usepackage{makecell}
\usepackage{tcolorbox}
\usepackage{pifont}
\usepackage{siunitx}
\usepackage{etoolbox}
\usepackage{nicematrix}
\usepackage{fontawesome5}
\usepackage{url}
\usepackage{hyperref}
\hypersetup{
    colorlinks=true,
    urlcolor=magenta   
}

\newcommand{\cmark}{{\color{green!70!black}\ding{52}}}
\newcommand{\xmark}{{\color{red!80!black}\ding{56}}}

\title{AudioRouter: Data Efficient Audio Understanding via \\RL based Dual Reasoning}

\author{
\textbf{Liyang Chen}$^{1,3}$ \quad
\textbf{Hongkai Chen}$^{3}$ \quad
\textbf{Yujun Cai}$^{2}$ \\
\textbf{Sifan Li}$^{3,4}$ \quad
\textbf{Qingwen Ye}$^{3}$ \quad
\textbf{Yiwei Wang}$^{4}$ \\[6pt]
$^{1}$University of California, Los Angeles \quad
$^{2}$The University of Queensland \\
$^{3}$vivo Mobile Communication Co., Ltd. \quad
$^{4}$University of California, Merced \\[4pt]
\texttt{\small Project Page: \url{https://liyang-chen-ucla.github.io/AudioRouter-page/}}
}

\begin{document}
\maketitle
\begin{abstract}
Large Audio Language Models (LALMs) have demonstrated strong capabilities in audio understanding and reasoning. However, their performance on fine grained auditory perception remains unreliable, and existing approaches largely rely on data intensive training to internalize perceptual abilities. We propose AudioRouter, a reinforcement learning framework that enables LALMs to improve audio understanding by learning when and how to use external audio tools. Rather than tightly coupling tool usage with audio reasoning, AudioRouter formulates tool use as an explicit decision making problem and optimizes a lightweight routing policy while keeping the underlying reasoning model frozen. Experimental results show that AudioRouter achieves substantial improvements on standard audio understanding benchmarks while requiring up to 600$\times$ less training data to learn tool usage compared with conventional training paradigms. These findings suggest that learning effective tool usage offers a data efficient and scalable alternative to internalizing perceptual abilities in LALMs.
\end{abstract}

\begin{figure*}[t]
  \centering
  \includegraphics[width=1.0\linewidth]{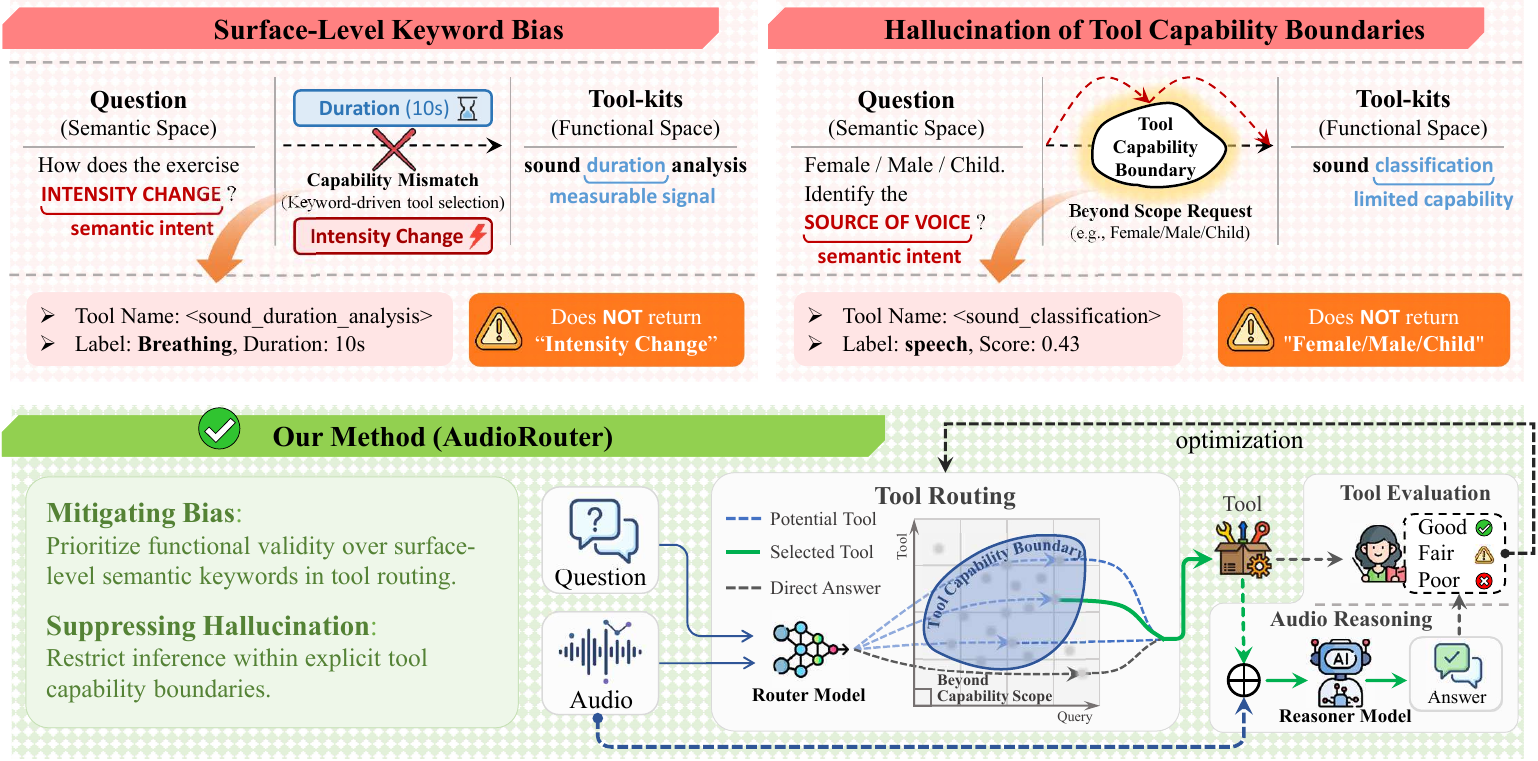}
  \vspace{-2mm}
    \caption{
    Two pervasive failure modes in tool-use systems and our capability-aware routing framework.
    \textbf{Top:} \emph{Surface-Level Keyword Bias} selects tools based on semantic keyword overlap despite functional mismatch, producing valid but irrelevant outputs (left).
    \emph{Hallucination of Tool Capability Boundaries} occurs when models extrapolate tool outputs beyond their explicit capability scope (right).
    \textbf{Bottom:} Our \emph{AudioRouter} models tool use as a capability-aware routing problem, prioritizing functional validity over surface cues and restricting inference within tool capability boundaries via relative reward.
    }

  \label{fig:performance_efficiency}
\end{figure*}

\section{Introduction}

Large Audio Language Models (LALMs) have recently demonstrated strong capabilities in semantic understanding, multimodal reasoning, and audio question answering. Despite this progress, their performance on low level auditory perception tasks, such as pitch estimation, temporal structure analysis, and event counting, remains unreliable. The dominant approach to addressing these limitations relies on large scale end-to-end optimization over audio CoT answer or audio answer data, where increasingly large collections of annotated audio text pairs are used to internalize perceptual competence and improve answer accuracy. In practice, this strategy faces substantial challenges: high quality audio annotation is costly, and perceptual abilities do not always scale proportionally with data volume.

From a system perspective, a key limitation of current LALMs lies in their difficulty acquiring fine grained and verifiable perceptual skills. At the same time, a wide range of mature and accurate audio processing tools already exist for low level tasks such as pitch tracking, event detection, and temporal analysis. This observation motivates a fundamental question: must audio language models continue to strengthen perceptual abilities through end-to-end training, or can they instead achieve reliable audio reasoning by reinforcing their ability to appropriately use external tools? However, making tool-use decisions directly within LALMs is itself challenging. As illustrated in Fig.~\ref{fig:performance_efficiency}, current models often exhibit \emph{surface level keyword bias}: tool routing can be driven by superficial semantic keywords rather than the tool’s functional validity, leading to capability mismatches. Moreover, models may \emph{hallucinate tool capability boundaries}, invoking tools beyond their supported scope and expecting outputs they cannot provide. These failure modes indicate that naïvely relying on the model to route tools end to end does not yield stable or reliable tool invocation behavior.

To address these challenges, we propose \textbf{AudioRouter}, a reinforcement learning based framework for tool use in LALMs. We explicitly formulate tool usage as a discrete decision making problem, where direct reasoning and tool-augmented reasoning are treated as alternative executable actions. A dedicated routing module, referred to as the Router, is introduced to decide whether a tool should be invoked and which tool to select, as illustrated in Fig.~\ref{fig:performance_efficiency}. Unlike prior end-to-end approaches, our framework decouples tool-use decisions from audio reasoning: the audio reasoning model is kept frozen, and only the routing policy is optimized, preventing interference with the model’s existing reasoning and perceptual representations.

Instead of relying on explicit supervision for tool selection, the Router is trained using reinforcement learning based on relative outcomes. Specifically, the learning signal is derived by comparing the final task performance achieved with and without tool usage, allowing the model to assess whether invoking a tool provides a tangible benefit. This outcome based objective enables the Router to learn effective tool-use strategies without requiring manually annotated tool-use labels.

Experimental results demonstrate that the proposed AudioRouter framework is highly data efficient and effective. Compared to mainstream end-to-end training paradigms that focus on internalizing perceptual abilities, our approach reduces the amount of training data required for learning tool usage by up to \textbf{600$\times$}, while simultaneously improving overall performance on multiple standard audio reasoning benchmarks. In several settings, our method achieves state of the art results. These findings suggest that learning how to use tools, rather than continuing to internalize perceptual abilities via end-to-end optimization, offers a more efficient and scalable pathway for advancing LALMs.

\section{Related Work}

\paragraph{LALM and Audio Reasoning}
LALMs have rapidly advanced in recent years, extending large language models to auditory perception and multimodal reasoning.
Early works such as AudioBERT and Speech2Text Transformers focused on aligning speech or acoustic representations with textual models.
More recent systems including Whisper LLM~\cite{radford2023whisper}, Qwen Audio~\cite{Qwen2-Audio}, and other Omni style architectures~\cite{Qwen2.5-Omni} integrate large scale audio encoders with instruction following LLMs to perform general purpose audio understanding, captioning, and conversational tasks.
Despite these improvements, LALMs continue to rely heavily on large curated datasets and struggle with low level auditory perception tasks such as pitch estimation, onset detection, and event counting.
Multi task benchmarks (e.g., MMAR~\cite{ma2025mmar}) further show that end-to-end LALMs remain limited in fine grained acoustic reasoning, even as they perform strongly on semantic or high level tasks.

\paragraph{Reinforcement Learning for LLM Reasoning}
Reinforcement learning (RL) has become a key mechanism for improving LLM reasoning, decision making, and planning.
Foundational methods such as RLHF~\cite{ouyang2022training}, RLAIF~\cite{lee2023rlaif}, and GRPO~\cite{shao2024deepseekmath} optimize language models through preference based rewards or structured advantages,
while more recent frameworks such as ReAct~\cite{yao2022react}, Reflexion~\cite{shinn2023reflexion}, and Tree of Thought~\cite{yao2023tree} extend RL to multi-step reasoning through action observation loops and self correction.
RL has also been applied to agent systems, program synthesis, embodied decision making, and multimodal control, where LLMs must select and execute sequences of actions.
These works demonstrate that RL can significantly improve coherence, task success rate, and strategic reasoning.
However, most RL-based reasoning frameworks focus on textual or visual domains;
relatively few explore RL for auditory tasks or for managing perceptual tools where action validity, tool selection, and trajectory consistency are essential components.

\paragraph{Tool use in LLM}
Tool-augmented language models have emerged as an important paradigm in enabling LLMs to access external computation, perception, and knowledge resources.
Representative works include Toolformer~\cite{schick2023toolformer}, ReAct~\cite{yao2022react}, HuggingGPT~\cite{shen2023hugginggpt}, Gorilla~\cite{patil2024gorilla}, and various API-augmented agent frameworks, which allow LLMs to call external search engines, calculators, code interpreters, or perception modules to solve complex tasks beyond the capability of purely parametric reasoning.
In multimodal contexts, systems such as MM-ReAct~\cite{yang2023mm}, ViperGPT~\cite{suris2023vipergpt}, and chain-of-visual-thought frameworks~\cite{huang2025vchain} demonstrate that LLMs can orchestrate perception modules to enhance visual reasoning.

While the audio domain benefits from a mature ecosystem of signal processing tools, existing LLMs rarely exhibit stable or systematic use of such tools.
Recent works such as AudioToolAgent~\cite{wijngaard2025audiotoolagent} and Audio-Maestro~\cite{lee2025audio} begin to introduce tool-augmented frameworks for LALMs, enabling closed source or black box audio models to interact with external audio tools.
However, these approaches largely rely on heuristic or scripted coordination, and robust learning based frameworks, particularly those driven by reinforcement learning, remain underexplored.

\begin{figure*}[t]
  \centering
  \includegraphics[width=1.0\linewidth]{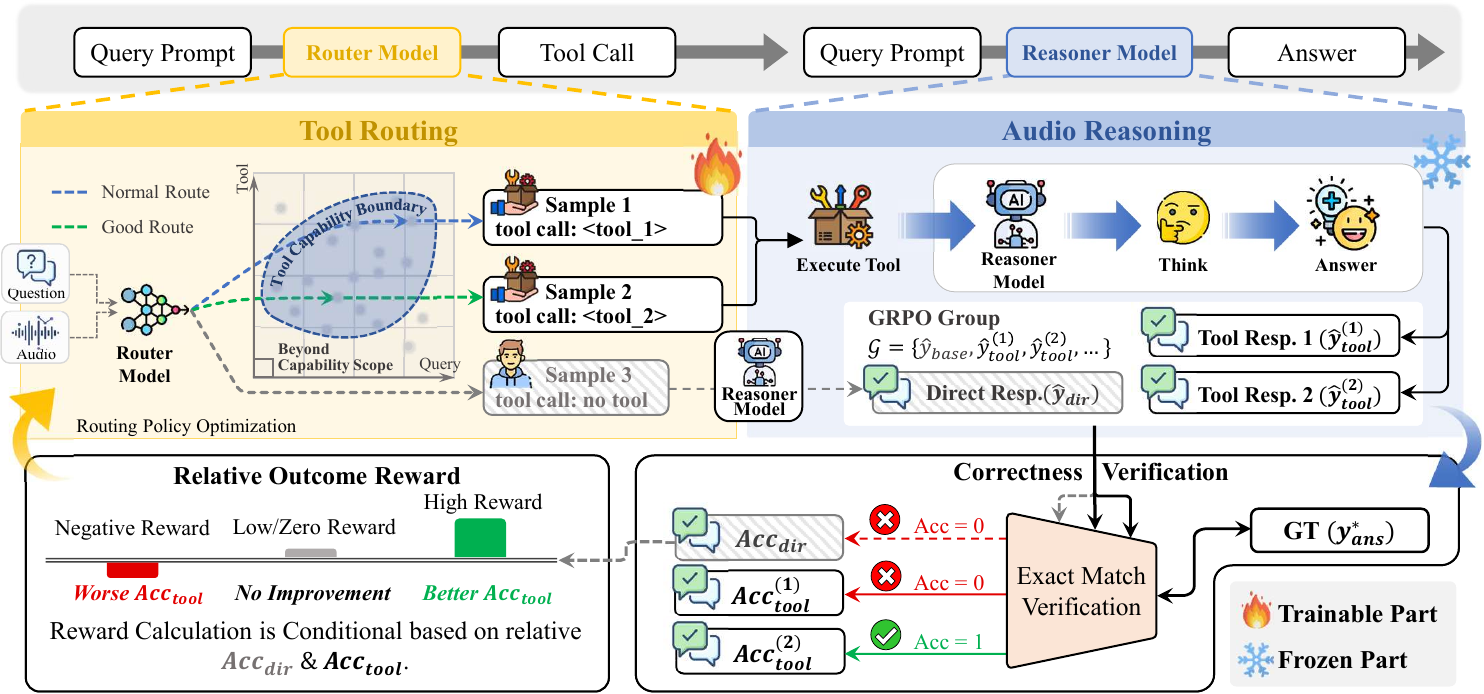}
  \vspace{-2mm}
    \caption{
    Overview of the proposed \emph{AudioRouter} framework.
    A Router first decides whether to invoke a tool based on capability aware routing, followed by a fixed Reasoner for audio reasoning.
    The Router is optimized via a \emph{relative outcome reward} by comparing tool-augmented and direct reasoning results, encouraging beneficial tool usage while suppressing redundant or harmful tool calls.
    }

  \label{fig:framework}
  \vspace{-3mm}
\end{figure*}

\section{Methodology}
\label{sec:method}

\subsection{Problem Formulation}
\label{sec:problem_formulation}

We define the tool-augmented audio question answering task as a decision making process under uncertainty. Given an input audio signal $x$, a natural language question $q$, and a set of candidate options $C = \{c_1, c_2, \dots, c_n\}$, the objective is to select the correct option $y^* \in C$.

Unlike conventional end-to-end LALMs, our system introduces a \textbf{Router} to determine the optimal reasoning path. We define the action space $\mathcal{A}$ as:
\begin{equation}
    a \in \mathcal{A} = \{ \text{Direct} \} \cup \{ t_1, t_2, \dots, t_K \}
\end{equation}
where $\text{Direct}$ denotes performing reasoning without external assistance, and $t_k \in \mathcal{T}$ represents the invocation of the $k$-th specialized audio tool.

The system's inference process is bifurcated based on the routing decision $a$:

\textbf{Direct Path ($a = \text{Direct}$):} The Reasoner $f_{\theta}$ predicts the answer directly from the raw inputs:
\begin{equation}
    \hat{y}_{\text{dir}} = f_{\theta}(x, q, C)
\end{equation}

\textbf{Tool-Augmented Path ($a = t_k$):} The system first invokes tool $t_k$ to extract structured perceptual evidence $r_k = t_k(x)$. The Reasoner then predicts the answer conditioned on this additional context:
\begin{equation}
    \hat{y}_{t_k} = f_{\theta}(x, q, C, r_k)
\end{equation}


\subsection{AudioRouter System}

To optimize the relative tool-use objective defined in the previous section, we design a reinforcement learning system that isolates tool-use decisions from answer reasoning. The system consists of three components: a routing policy $\pi(a \mid x, q, C)$, a set of external audio tools $\mathcal{T} = \{t_1,\dots,t_K\}$, and a frozen audio reasoning model $f_\theta$. An illustration of the overall framework is shown in Fig.~\ref{fig:framework}.
This decoupled design is critical for data efficiency.
The Router does not learn audio perception or answer reasoning; instead, it functions purely as a lightweight task dispatcher that selects among a small set of discrete actions.
Consequently, the learning problem faced by the Router has substantially lower complexity than end-to-end optimization of perceptual or reasoning models.
By restricting learning to a low dimensional dispatch policy while keeping the reasoning backbone frozen, AudioRouter significantly reduces sample complexity and enables effective training from a limited number of tool-use examples.

\subsubsection{Tool Router Model}

The router implements a policy $\pi(a \mid x, q, C)$ that determines whether to perform direct reasoning or invoke an external audio tool. Given a question $q$, optional audio representation $x$, and candidate options $C$, the router outputs an action:
\begin{equation}
a \in \mathcal{A} = \{\text{Direct}\} \cup \{t_1, t_2, \dots, t_K\}.
\end{equation}

The router can operate in two configurations: (i) a \textbf{multimodal setting}, where audio derived features $x$ are provided; or (ii) a \textbf{text only setting}, where decisions are conditioned solely on $q$. Both configurations share the same action space and optimization objective, and are evaluated in our experiments.

When the router selects a tool action $a = t_k$, it outputs a syntactically valid tool call:
\begin{center}
\texttt{<tool\_call> $t_k$ </tool\_call>}
\end{center}
where $t_k \in \mathcal{T}$ is a predefined and executable tool. This constraint ensures that all router actions can be reliably executed during inference. The router is trained to learn a precise decision policy: invoking tools only when they are expected to improve the final reasoning outcome, selecting appropriate tools when necessary, and suppressing unnecessary or harmful tool calls.

\subsubsection{Audio Reasoner Model}

We adopt a strong pretrained audio language model as the reasoning backbone and keep it frozen throughout training. Given the question $q$, audio input $x$, candidate options $C$, and optional tool output $r$, the reasoning model produces the final prediction:
\begin{equation}
\hat{y} = f_\theta(x, q, C, r),
\end{equation}
where $r = \emptyset$ in the direct reasoning case and $r = r_k = t_k(x)$ when a tool $t_k$ is invoked. Accordingly, the reasoning model produces the two outcomes
$\hat{y}_{\text{dir}}$ and $\hat{y}_{t_k}$ as defined in
Section~\ref{sec:problem_formulation}.

By keeping $f_\theta$ fixed, differences between $\hat{y}_{\text{dir}}$ and $\hat{y}_{t_k}$ can be attributed solely to the effect of tool usage. This property is essential for evaluating the relative utility of tool invocations under identical reasoning capacity.

\subsubsection{Relative Outcome Reward}
A central challenge in learning tool-use policies lies in defining an appropriate learning signal.
Naïve absolute rewards that evaluate tool-augmented predictions in isolation are fundamentally misaligned with our objective: tools are not the goal themselves, but merely optional means to improve reasoning.
Optimizing such absolute rewards tends to encourage unnecessary or excessive tool usage, as the router may learn to invoke tools even when direct reasoning already suffices.

To avoid these failure modes, we adopt a \textbf{Relative Outcome Reward}.
Rather than judging a tool invocation in isolation, the Relative Outcome Reward explicitly compares reasoning outcomes with and without tool assistance under identical reasoning capacity.
This comparison directly answers the only question that matters for tool routing: \emph{does invoking this tool provide a genuine marginal benefit over direct reasoning?}
As a result, relative outcome comparison constitutes the only learning signal that is strictly aligned with our routing objective.

\begin{figure}[t]
  \centering
  \includegraphics[width=1.0\linewidth]{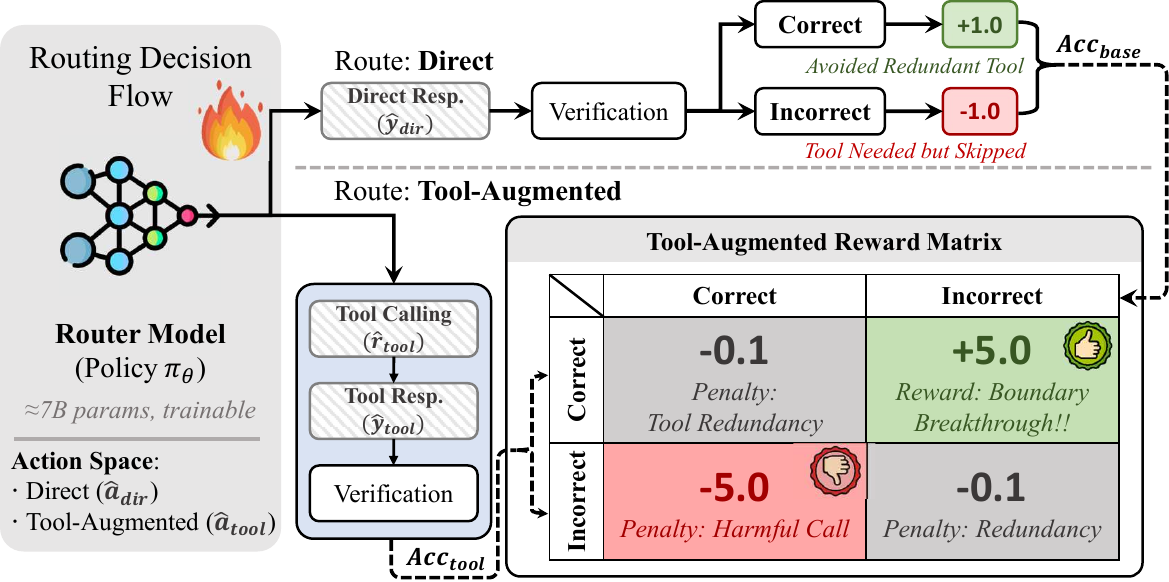}
  \vspace{-2mm}
  \caption{
    Relative outcome reward for training the Router policy.
    Tool-augmented decisions are rewarded or penalized based on their correctness \emph{relative} to direct reasoning, encouraging boundary breaking tool usage while suppressing redundant or harmful tool calls.
    }

  \label{fig:relative_reward}
\end{figure}

For a given input $(x, q, C)$, let $\hat{y}_{\text{dir}}$ denote the prediction produced when the router selects $a=\text{Direct}$, and let $\hat{y}_{t_k}$ denote the prediction produced when the router selects $a=t_k$ and the corresponding tool output is provided to the reasoning model. The reward assigns a scalar value based on the relative correctness of these two outcomes, reflecting whether the tool invocation is beneficial, redundant, or harmful (see Fig.~\ref{fig:relative_reward}).

Specifically, the relative outcome reward is defined based on the routing action $a$ and the correctness of the resulting predictions. Let $\text{acc}_{\text{dir}} \in \{0,1\}$ and $\text{acc}_{t_k} \in \{0,1\}$ denote the answer correctness indicators obtained by matching the direct prediction $\hat{y}_{\text{dir}}$ and the tool-augmented prediction $\hat{y}_{t_k}$ with the ground truth answer $y^*$, respectively.

\paragraph{Tool Action ($a = t_k$).}
When the router invokes a tool $t_k$, the reward is defined by comparing the correctness of the tool-augmented outcome against the direct baseline:
\begin{equation}
R_{tool} =
\begin{cases}
+ 5.0, & \text{acc}_{t_k} = 1 \wedge \text{acc}_{\text{dir}} = 0, \\
- 5.0, & \text{acc}_{t_k} = 0 \wedge \text{acc}_{\text{dir}} = 1, \\
- 0.1, & \text{acc}_{t_k} = \text{acc}_{\text{dir}}, \\
0, & \text{otherwise}.
\end{cases}
\end{equation}

\paragraph{Direct Action ($a = \text{Direct}$).}
When the router selects the direct action, the reward depends solely on the correctness of the direct prediction:
\begin{equation}
R_{dir} =
\begin{cases}
+ 1.0, & \text{acc}_{\text{dir}} = 1, \\
- 1.0, & \text{acc}_{\text{dir}} = 0.
\end{cases}
\end{equation}

By optimizing this relative outcome reward, the router learns to invoke tools only when they yield a positive correctness gain over direct reasoning, while suppressing redundant or harmful tool usage. As demonstrated in our experiments, this reward formulation is essential for learning robust and high precision routing policies under noisy and heterogeneous tool outputs.

\begin{figure*}[t]
    \centering
    \includegraphics[width=\linewidth]{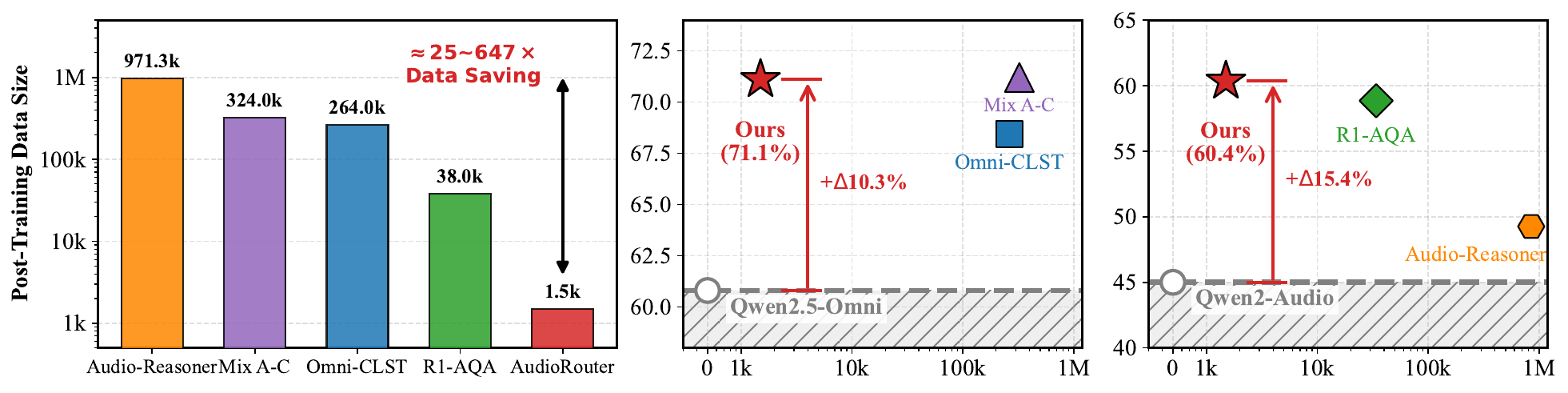}
    \caption{Main results on MMAU-mini and MMAR.
    \textbf{Left:} Post training data scale comparison.
    AudioRouter achieves $25\times$–$647\times$ data savings.
    \textbf{Right:} Performance comparison under Qwen2.5-Omni and Qwen2-Audio backbones,
    where AudioRouter consistently improves accuracy over end-to-end baselines
    and remains competitive with closed source models.}

    \label{fig:main-result}
\end{figure*}

\section{Experiment}

\subsection{Experimental Setup}

\paragraph{Models.}
We compare end-to-end audio language models with our proposed \textbf{AudioRouter}.
We use two baselines with different capacities:
\textbf{R1-AQA}~\cite{li2025reinforcement}, trained on Qwen2-Audio-7B~\cite{Qwen2-Audio},
and \textbf{Omni-CLST}~\cite{zhao2025omni}, trained on Qwen2.5-Omni-7B~\cite{Qwen2.5-Omni}.
Both baselines perform direct audio reasoning without tool routing.

For AudioRouter, the corresponding backbone is used as a frozen reasoner for a fair comparison.
The router is implemented using Qwen2.5-Omni-7B~\cite{Qwen2.5-Omni} and is the only trainable component.
A separate router is trained for each reasoner backbone.

\paragraph{Training Data.}
We train the router using a lightweight set of 1.5k audio QA samples derived from the MMAU Pro dataset ~\cite{kumar2025mmau}.
Following standard filtering practices, we retain only samples with short audio clips and multiple choice questions, and exclude all open-ended questions to ensure reliable and unambiguous reward evaluation.

\begin{table}[h]
    \centering
    \scriptsize 
    \renewcommand{\arraystretch}{1.20} 
    \setlength{\tabcolsep}{3pt} 
    \begin{tabular}{p{0.4\linewidth} p{0.55\linewidth}}
        \hline
        \textbf{Module} & \textbf{Underlying Model / Library} \\
        \hline
        Speech Recognition          & Whisper-large-v3~\cite{radford2023whisper} \\
        Sound Classification        & AST~\cite{gong2021ast} \\
        Sound Duration Analysis     & AST (sliding window)~\cite{gong2021ast} \\
        Sound Temporal Analysis     & AST (sliding window)~\cite{gong2021ast} \\
        Chord Recognition           & autochord~\cite{bayron2021autochord} \\
        Audio Feature Extraction    & librosa~\cite{mcfee2015librosa} \\
        \hline
    \end{tabular}
    \caption{\textbf{Audio tool modules and underlying models/libraries used in our system.}}
    \label{tab:audio_tools}
\end{table}

\paragraph{Tool-kits Usage.}
All external audio tools follow a unified JSON based API that returns 
structured predictions.  
The router integrates these predictions when deciding whether tool assistance 
is beneficial.  
Full tool specifications are listed in Table~\ref{tab:audio_tools}.

\paragraph{Reinforcement Learning Details.}
The router is finetuned using GRPO with single step tool-use rollouts.
We use a learning rate of $5 \times 10^{-5}$ and a KL regularization coefficient of $\beta = 0.01$.
Parameter efficient finetuning is performed using LoRA with rank $r = 16$, scaling factor $\alpha = 32$, and dropout probability $p = 0.05$.
Training is conducted for one epoch over the 1.5k training samples.

Prior to GRPO optimization, we perform a brief warm-up stage using only the format reward to stabilize the router’s output structure.

\paragraph{Evaluation Benchmarks.}
We evaluate our method on two standard multi task audio reasoning benchmarks:  
\textbf{MMAU-mini} ~\cite{sakshi2024mmau} and \textbf{MMAR} ~\cite{ma2025mmar}, which together cover a wide range of 
low level auditory perception and high level multimodal reasoning tasks.

\definecolor{highlightblue}{RGB}{235, 245, 255} 
\definecolor{headergray}{RGB}{240, 240, 240}    

\begin{table}[h]
    \setlength{\tabcolsep}{4pt} 
    \centering
    \caption{Performance on MMAU-mini benchmark.}
    \label{tab:mmau-results}
    \small
    
    \begin{NiceTabular}{lcccc}[colortbl-like] 
        \toprule
        \rowcolor{headergray} 
        \textbf{Model} & \textbf{Sound} & \textbf{Music} & \textbf{Speech} & \textbf{Avg} \\
        \midrule
        Random Guess & 26.7 & 24.6 & 26.7 & 26.0 \\
        \midrule
        
        \multicolumn{5}{l}{\textit{Qwen2-Audio Based}} \\
        \addlinespace[0.2em]
        \quad Qwen2-Audio & 51.2 & 58.6 & 69.0 & 59.6 \\
        \quad R1-AQA       & 60.1 & 65.4 & \textbf{73.7} & 66.4 \\
        
        \rowcolor{highlightblue}
        $\rightarrow$ \textbf{AudioRouter} & \textbf{63.8} & \textbf{67.2} & 71.2 & \textbf{67.4} {\scriptsize\textcolor{ForestGreen}{(+1.0)}}\\

        \addlinespace[0.3em]
        \multicolumn{5}{l}{\textit{Qwen2.5-Omni Based}} \\ 
        \addlinespace[0.2em]
        \quad Qwen2.5-Omni & 62.4 & 68.1 & 78.0 & 69.5 \\
        \quad Omni-CLST    & 68.5 & 72.4 & 80.8 & 73.9 \\
        \rowcolor{highlightblue}
        $\rightarrow$ \textbf{AudioRouter} & \textbf{74.2} & \textbf{76.9} & \textbf{82.3} & \textbf{77.8} {\scriptsize\textcolor{ForestGreen}{(+3.9)}}\\
        \bottomrule
    \end{NiceTabular}
\end{table}

\begin{table*}[t]
\small
\centering
\caption{Performance on MMAR benchmark. \textbf{AudioRouter} (Ours) consistently outperforms representative baselines. Bold numbers indicate the best results.}
\label{tab:mmar-results}

    \begin{NiceTabular}{c ccc cccc c}
    \toprule
    \rowcolor{headergray} 
    \multirow{2}{*}{\textbf{Model}} & \multicolumn{3}{c}{\textbf{Single Modality (\%)}} & \multicolumn{4}{c}{\textbf{Mixed Modalities (\%)}} & \multirow{2}{*}{\textbf{Avg (\%)}} \\
    \cmidrule(lr){2-4} \cmidrule(lr){5-8}
    \rowcolor{headergray}
    & Sound & Music & Speech & So-Mu & So-Sp & Mu-Sp & So-Mu-Sp & \\
    \midrule
    Random Guess & 29.4 & 25.9 & 31.5 & 25.0 & 29.3 & 31.1 & 28.1 & 29.3 \\
    \midrule
    \multicolumn{9}{c}{\textit{Qwen2-Audio Based}} \\ 
    \midrule
    Qwen2-Audio & 28.5 & 30.2 & 35.4 & 26.8 & 31.0 & 32.5 & 28.4 & 30.4 \\ \addlinespace[0.5ex]
    R1-AQA & 46.2 & 54.1 & 66.8 & 43.5 & 50.2 & 53.7 & 44.6 & 51.3 \\ \addlinespace[0.5ex]
    \rowcolor{highlightblue} 
    $\rightarrow$ \textbf{AudioRouter} & 
    \textbf{46.5} {\scriptsize\textcolor{ForestGreen}{(+0.3)}} & 
    \textbf{54.6} {\scriptsize\textcolor{ForestGreen}{(+0.5)}} & 
    \textbf{67.1} {\scriptsize\textcolor{ForestGreen}{(+0.3)}} & 
    \textbf{44.1} {\scriptsize\textcolor{ForestGreen}{(+0.6)}} & 
    \textbf{50.9} {\scriptsize\textcolor{ForestGreen}{(+0.7)}} & 
    \textbf{54.3} {\scriptsize\textcolor{ForestGreen}{(+0.6)}} & 
    \textbf{45.4} {\scriptsize\textcolor{ForestGreen}{(+0.8)}} & 
    \textbf{51.8} {\scriptsize\textcolor{ForestGreen}{(+0.5)}} \\
    \midrule
    \multicolumn{9}{c}{\textit{Qwen2.5-Omni Based}} \\ 
    \midrule
    Qwen2.5-Omni  & 45.6 & 55.2 & 68.4 & 42.1 & 51.5 & 55.8 & 46.1 & 52.1 \\ \addlinespace[0.5ex]
    Omni-CLST & 58.2 & 64.7 & 76.5 & 54.9 & 63.2 & 67.4 & \textbf{56.1} & 63.0 \\ \addlinespace[0.5ex]
    \rowcolor{highlightblue}
    $\rightarrow$ \textbf{AudioRouter} & 
    \textbf{61.4} {\scriptsize\textcolor{ForestGreen}{(+3.2)}} & 
    \textbf{66.1} {\scriptsize\textcolor{ForestGreen}{(+1.4)}} & 
    \textbf{77.2} {\scriptsize\textcolor{ForestGreen}{(+0.7)}} & 
    \textbf{57.2} {\scriptsize\textcolor{ForestGreen}{(+2.3)}} & 
    \textbf{65.1} {\scriptsize\textcolor{ForestGreen}{(+1.9)}} & 
    \textbf{68.3} {\scriptsize\textcolor{ForestGreen}{(+0.9)}} & 
    55.5 {\scriptsize\textcolor{red}{(-0.6)}} & 
    \textbf{64.4} {\scriptsize\textcolor{ForestGreen}{(+1.4)}} \\
    \bottomrule
    \end{NiceTabular}
\end{table*}

\subsection{Main Results}

\paragraph{Overall Performance Comparison.}
We evaluate the proposed AudioRouter method on two benchmarks, MMAU-mini and MMAR, and compare it with end-to-end audio reasoning models (see Fig.~\ref{fig:main-result}, Tab.~\ref{tab:mmau-results}, and Tab.~\ref{tab:mmar-results}). The results demonstrate that AudioRouter consistently yields performance improvements across both backbone settings, achieving state-of-the-art results.

\paragraph{Data Efficiency.}
The left side of Fig.~\ref{fig:main-result} illustrates the post training data scales required by different methods. AudioRouter utilizes only 1.5k training samples, significantly less than the 38.0k to 971.3k samples required by competing methods. This represents a data efficiency advantage of approximately 25$\times$ to 647$\times$. Notably, AudioRouter consistently improves upon end-to-end baselines even with minimal data, suggesting that learning an optimal routing policy is a highly cost effective way to enhance performance without increasing model parameters.

\paragraph{Analysis on MMAU-mini.}
On the MMAU-mini benchmark (Tab.~\ref{tab:mmau-results}), AudioRouter shows robust adaptability across different architectures. Specifically, when using the Qwen2-Audio series backbone, the baseline R1-AQA achieves an average accuracy of 66.4\%, which AudioRouter elevates to 67.4\%. This improvement is primarily driven by the Sound and Music subcategories, although a marginal decline is observed in Speech. For the Qwen2.5-Omni series, the gains are even more pronounced; AudioRouter raises the average accuracy from 73.9\% to 77.8\%, demonstrating consistent improvements across all categories, including Sound, Music, and Speech.

\paragraph{Analysis on MMAR.}
The MMAR benchmark (Tab.~\ref{tab:mmar-results}) further confirms the efficacy of the routing strategy. For the Qwen2-Audio series, AudioRouter increases the average accuracy from 51.3\% to 51.8\%, with small but consistent gains in both single modality and mixed modality subsets. Similarly, for the Qwen2.5-Omni series, the average accuracy improves from 63.0\% to 64.4\%. These gains are largely attributed to the Sound category and various mixed modality settings. While there is a slight performance dip in the So--Mu--Sp subset, the overall trend remains positive across the backbone.

\paragraph{External Reference.}
Furthermore, Fig.~\ref{fig:main-result} compares AudioRouter with several closed-source models, such as Audio Reasoner ~\cite{xie2025audio} and Mix A-C ~\cite{he2025measuring}. AudioRouter exhibits competitive performance across multiple settings, providing a robust reference for its overall effectiveness and high precision routing capabilities.

\subsection{Effectiveness of AudioRouter}

To evaluate the effectiveness of the proposed AudioRouter approach, we conduct controlled comparisons on MMAU-mini and MMAR (Table~\ref{tab:tool-rl-results}). 
The compared methods cover both end-to-end modeling paradigms and tool-augmented reasoning under the Router--Reasoner framework. 
Specifically, we include a base audio language model without tool routing, a strong task adapted end-to-end baseline, an end-to-end tool-use reinforcement learning variant, and our proposed AudioRouter method.

\paragraph{Tool-use RL}
The Tool-use RL variant adopts a Router--Reasoner style inference pipeline but does not explicitly introduce a separate routing module. 
Instead, reinforcement learning is applied end-to-end to enhance the model’s tool-use capability, allowing it to implicitly learn when to invoke tools and how to incorporate tool outputs during answer generation. 
As shown in Table~\ref{tab:tool-rl-results}, this strategy improves performance over the base Qwen2.5-Omni model on both MMAU-mini and MMAR, but remains consistently below the strong end-to-end baseline Omni-CLST.

\paragraph{AudioRouter}
In contrast, AudioRouter explicitly introduces a lightweight routing policy and applies reinforcement learning only to the router, while keeping the audio reasoning backbone frozen throughout training. 
This design decouples tool-use decisions from answer generation, enabling the routing policy to be optimized via outcome based reinforcement signals without altering the pretrained perceptual and reasoning capabilities of the backbone model. 
As reported in Table~\ref{tab:tool-rl-results}, AudioRouter achieves the strongest performance on both benchmarks, outperforming Omni-CLST by 3.9 and 1.4 points on MMAU-mini and MMAR, respectively, and consistently surpassing the end-to-end Tool-use RL variant.

\begin{table}[h]
\centering
\small
\caption{Ablation on Tool-use RL vs.\ our AudioRouter. Improvements over Omni-CLST are shown in green.}
\label{tab:tool-rl-results}
\setlength{\tabcolsep}{4pt} 
\begin{NiceTabular}{lccc}
    \toprule
    \rowcolor{headergray}
    \textbf{Model} & \textbf{MMAU-mini} & \textbf{MMAR} & \textbf{Avg.} \\
    \midrule
    Qwen2.5-Omni & 69.5 & 52.1 & 60.8 \\
    Omni-CLST & 73.9 & 63.0 & 68.5 \\
    \midrule
    Tool-use RL & 71.4 & 53.4 & 62.40 \\
    \rowcolor{highlightblue}
    $\rightarrow$ \textbf{AudioRouter} & 
    \textbf{77.8} {\scriptsize\textcolor{ForestGreen}{(+3.9)}} & 
    \textbf{64.4} {\scriptsize\textcolor{ForestGreen}{(+1.4)}} & 
    \textbf{71.10} {\scriptsize\textcolor{ForestGreen}{(+2.6)}}\\
    \bottomrule
\end{NiceTabular}
\end{table}

\subsection{Router Evaluation}

To further assess the quality of routing decisions, we evaluate different routing strategies on MMAU-mini and MMAR using a controlled evaluation protocol (Table~\ref{tab:router-eval-results}). 
This evaluation focuses on the effectiveness of the routing policy itself, rather than the overall model capacity.

We compare our RL-trained router with a random router, a router based on Qwen2.5-Omni, and a router based on Gemini-2.5-flash. 
As shown in Table~\ref{tab:router-eval-results}, random routing yields the lowest performance, while model-based routers achieve noticeably better results, reflecting their ability to make more informed routing decisions.

Across both benchmarks, AudioRouter consistently achieves the highest accuracy, outperforming all baseline routers. 
These results indicate that reinforcement learning enables the router to learn more effective routing strategies, leading to improved overall reasoning performance.

\begin{table}[t]
\centering
\small
\setlength{\tabcolsep}{6pt} 
\setlength{\tabcolsep}{4pt} 
\begin{NiceTabular}{lccc}
    \toprule
    \rowcolor{gray!10} 
    \textbf{Router Type} & \textbf{MMAU-mini} & \textbf{MMAR} & \textbf{Avg.} \\
    \midrule
    Random & 65.2 & 50.1 & 57.6 \\
    Qwen2.5-Omni & 72.4 & 60.5 & 66.4 \\
    Gemini-2.5-flash & 74.8 & 61.2 & 68.0 \\
    \midrule
    \rowcolor{highlightblue}
    $\rightarrow$ \textbf{AudioRouter} & 
    \textbf{77.8} {\scriptsize\textcolor{ForestGreen}{(+3.0)}} & 
    \textbf{64.4} {\scriptsize\textcolor{ForestGreen}{(+3.2)}} & 
    \textbf{71.1} {\scriptsize\textcolor{ForestGreen}{(+3.1)}} \\
    \bottomrule
\end{NiceTabular}
\caption{Router evaluation on MMAU-mini and MMAR.
Comparison of different routing strategies. AudioRouter achieves the highest accuracy on both benchmarks.}
\label{tab:router-eval-results}
\end{table}

\subsection{Case Study}

To qualitatively evaluate the effectiveness of AudioRouter, we analyze a representative failure case of the baseline Omni-CLST and show how our router mitigates it through targeted tool invocation.

\paragraph{Physical Parameter Anchoring.}
In this example, the baseline model makes an ungrounded high level classification, incorrectly predicting \textit{Bossa Nova}. When routed to the \texttt{genre\_analysis} tool, the system obtains concrete physical cues, including a \textbf{tempo} of 129.2 BPM and a \textbf{pitch} of 1520.58 Hz. 

These measurements provide decisive constraints: the fast tempo is characteristic of electronic or club oriented styles and effectively rules out Bossa Nova, which typically falls within 60--90 BPM. Guided by these physical anchors, the reasoning model correctly identifies the \textit{Nocturnal scene} electronic style, illustrating how tool based perceptual evidence can ground semantic reasoning more efficiently than end-to-end perception.

\definecolor{lightgray}{RGB}{245,245,245}
\definecolor{errorred}{RGB}{180,30,30}
\definecolor{highlightyellow}{RGB}{255, 250, 205}

\begin{figure}[h]
    \begin{tcolorbox}[
        colback=white,
        colframe=black,
        boxrule=0.6pt,
        arc=2pt,
        title=Case: \texttt{genre\_analysis},
        fonttitle=\bfseries\small,
        fontupper=\small,
        left=5pt, right=5pt, top=5pt, bottom=5pt,
        before skip=6pt,
        after skip=3pt
    ]
    
    \noindent\colorbox{gray!15}{%
    \parbox{\linewidth}{\footnotesize\faQuestionCircle\ \textbf{ Question}}}
    \vspace{3pt} \\
    What style of music is the accompaniment before the electronic diva appears?
    
    \medskip
    
    \noindent\colorbox{gray!15}{%
    \parbox{\linewidth}{\footnotesize\faListUl\ \textbf{ Choices}}}
    \vspace{3pt} \\
    {\small 
    Bossa Nova, \ 
    K pop, \ 
    \tcbox[
        on line,
        boxsep=0pt,
        left=2pt, right=2pt, top=1pt, bottom=1pt,
        colback=ForestGreen!10,
        colframe=ForestGreen!50,
        arc=2pt,
        boxrule=0.5pt
    ]{\textcolor{ForestGreen}{\textbf{\normalsize{Electronic Jazz}}}}
    }
    
    \tcblower
    
    \textbf{Direct Answer (w/o tools).}
    
    \begingroup
    \footnotesize\itshape\color{gray}
    “Based on an initial impression, the accompaniment sounds relaxed and rhythmic.
    \textcolor{errorred}{This leads to an incorrect association with Bossa Nova},
    without considering tempo or timbral characteristics.”
    \par\endgroup
    
    \smallskip
    $\rightarrow$ \textcolor{errorred}{\textbf{Bossa Nova}} \xmark
    
    \medskip
    \textbf{Tool Result.}
    
    \begin{tcolorbox}[
        colback=lightgray,
        colframe=black!20,
        boxrule=0.4pt,
        arc=1pt,
        left=4pt, right=4pt, top=3pt, bottom=3pt
    ]
    \ttfamily\footnotesize
    [
      {\{"tempo": 129.20\}},\\
      {\{"pitch": 1520.58\}},\\
      \{"timbre": 2716.82\}
    ]
    \end{tcolorbox}
    
    \medskip
    \textbf{Tool-Augmented Answer.}
    
    \begingroup
    \footnotesize\itshape\color{gray}
    “The tool analysis reports a tempo of 129.20 BPM, which is characteristic of
    electronic or club oriented genres and
    \colorbox{highlightyellow}{\color{black}excludes slower styles such as Bossa Nova}.
    The elevated pitch value (1520.58 Hz), together with the timbre features,
    supports the presence of a female vocal line layered over
    \colorbox{highlightyellow}{\color{black}synthetic and atmospheric accompaniment}.”
    \par\endgroup
    
    \smallskip
    $\rightarrow$ \textcolor{ForestGreen}{\textbf{Electronic Jazz}} \cmark
    
    \end{tcolorbox}
    {\footnotesize\emph{
    Note: Segments highlighted with
    \colorbox{highlightyellow}{\strut\hspace{0.8em}}
    indicate evidence where tool usage provides helpful information.
    }}
\end{figure}

\section{Conclusion}

This paper presented \textbf{AudioRouter}, a reinforcement learning based framework for learning tool usage in LALMs. By explicitly formulating tool usage as a decision making problem and decoupling tool-use decisions from audio reasoning, our approach enables effective and data efficient learning without modifying the underlying reasoning model. 

Experimental results demonstrate that AudioRouter improves overall audio reasoning performance while significantly reducing the amount of training data required for learning tool usage. These findings suggest that learning how to use external tools offers a practical and scalable alternative to end-to-end perceptual learning, and provides a promising direction for advancing tool-augmented audio language models.

\section{Limitations}

The relative outcome reward relies on a fixed reasoning model as a comparison baseline. While this design avoids explicit supervision on tool necessity and stabilizes training, the Router’s learning signal is inherently bounded by the capability of the underlying Reasoner.

In addition, our experiments focus on short-form, closed-set audio reasoning tasks with a limited set of audio tools. Although the identified failure modes and the AudioRouter framework are modality-agnostic, extending the approach to long-form reasoning, open-ended questions, and more diverse tool capabilities remains an important direction for future work.


\bibliography{custom}

\appendix


\end{document}